\documentstyle[prd,aps]{revtex}

\draft
\input epsf
\begin{document}
\title{ Improvements to the Method of Dispersion Relations\\
for $B$ Nonleptonic Decays}
\author{I. Caprini and L. Micu}
\address{National Institute of Physics and Nuclear Engineering
 POB MG 6, Bucharest, R-76900 Romania}
\author{C. Bourrely}
\address{Centre de Physique Th\'eorique
\footnote{Laboratoire propre au CNRS-UPR 7061}, CNRS-Luminy
Case 907, \\F-13288 Marseille Cedex 9 - France}
\maketitle
\begin{abstract}
We  bring some clarifications and improvements to the method of
dispersion relations in the external masses variables, that we proposed recently
for investigating the final state interactions in the $B$ nonleptonic decays.
We first present arguments for the existence of an additional term in the
dispersion representation, which arises from an equal-time commutator in the
LSZ formalism and can be approximated by the conventional factorized
amplitude. The reality properties of the spectral function and the
Goldberger-Treiman procedure to perform the hadronic unitarity sum are
analyzed in more detail. We also improve the treatment of the strong
interaction part by including the contributions of both $t$ and $u$-channel
trajectories in the Regge amplitudes. Applications to the $B^0\to \pi^+\pi^-$
and $B^+\to \pi^0 K^+$ decays are presented. 
\end{abstract}

\section{Introduction}

In a recent paper \cite{CaMi}, we discussed the rescattering effects in the
nonleptonic  $B$ decays into light pseudoscalar mesons, calculated by a method
of  dispersion relations in terms of the external masses. We recall that the
analytic continuation in the external masses was originally investigated 
in the frame of axiomatic field theory \cite{KaWi}. 
In \cite{CaMi} we used an approach
based on the Lehmann-Symanzik-Zimmermann (LSZ) reduction formalism \cite{LSZ},
and we showed that the weak amplitude satisfies a dispersion representation in
the mass squared of one final particle, with a spectral function given by the
hadronic unitarity sum  associated to rescattering effects. We mention that the
analyticity domain in this heuristic treatment is  much larger than the domain
obtained by rigorous techniques. 
In the present work we discuss in more detail some aspects of the dispersion
relations proposed in \cite{CaMi}.

\section{Discussion of the method}

Defining the  weak decay amplitude $A_{B\to P_1 P_2}=A(m_B^2, m_1^2, m_2^2)$,
where $P_1$, $P_2$ are light pseudoscalar mesons, we considered in \cite{CaMi} 
the LSZ reduction relation \cite{LSZ}, \cite{Bart}
\begin{equation}\label{lsz1}
A (m_B^2, k_1^2, m_2^2)= {i\over \sqrt{2 \omega_1}}\int {\mathrm d}x
 e^{ik_1 x} \theta (x_0) \langle P_2(k_2)|[\eta _1(x),
{\mathcal H}_w(0)]|B(p)\rangle\,,
\end{equation}
where $\eta_1(x)= {\cal K}_x \phi_1(x)$ denotes the source of the meson $P_1$.
This relation was  the starting point for showing that the weak amplitude
satisfies a dispersion representation in the mass variable $k_1^2$, with a
discontinuity  given by
\begin{equation}\label{spectr}
\mbox{Disc}\, A(m_B^2, k_1^2, m_2^2)= {1\over 2 \sqrt{2 \omega_1}}
\sum_{n}\delta  (k_1+k_2-p_n) \langle P_2(k_2)|
\eta _1|n\rangle\langle n| {\mathcal H}_w|B(p)\rangle\,.
\end{equation}  In each term of the sum  (\ref{spectr})
the first  matrix element represents the amplitude of the strong transition 
from the intermediate state $|n\rangle$ to the final state $P_1P_2$, 
for an  off-shell meson $P_1$ with an the invariant mass squared $k_1^2$,
multiplied by the amplitude of the weak transition of $B$ into
the same intermediate state. Therefore, the discontinuity  (\ref{spectr})
describes  the final state rescattering effects in the decay $B\to P_1 P_2$.
Note that all the particles involved in the weak transition  are real, on-shell
particles.

In Ref. \cite{CaMi} we obtained the whole amplitude from its discontinuity by
means of a dispersion relation without subtractions. We note however that the
presence of an additional term in the dispersion relations is not excluded,
even if the subtractions in the dispersion integral are not necessary. To see
the origin of such additional terms we notice that in deriving the relation
({\ref{lsz1}) we retained only the contribution given by the action of the
Klein-Gordon operator  ${\cal K}_x$ on the interpolating field $\phi_1(x)$,
since this was of interest for the analytic continuation. However, in the LSZ
formula there are some additional terms produced by the action of  ${\cal K}_x$
upon the function $\theta(x_0)$. These terms can be written as a sum of equal
time commutators in the form \cite{Bart}
\begin{equation}\label{degen}
 {i\over \sqrt{2 \omega_1}}\int {\mathrm d}x
 e^{ik_1 x} \delta(x_0) \langle P_2(k_2)|-i k_{10}[\phi_1(x),
{\mathcal H}_w(0)]\,+\, [\partial_0\phi_1(x),
{\mathcal H}_w(0)|B(p)\rangle\,.
\end{equation}
As discussed in \cite{Bart}, these terms can contribute only if there is a
``direct'' connection between the interpolating field  $\phi_1(x)$ and the
operator
${\mathcal H}_w(0)$. Let us suppose that we applied the LSZ reduction 
formula to the
final meson which does not contain the spectator quark in the $B$ decay. 
Then we can assume that one of the currents entering the expression of 
${\mathcal H}_w(0)$, more exactly the current which contains the fields of 
the quarks going into the final meson $P_1$, 
is related to the interpolating field by
$j^{(1)}_\mu \approx \partial_\mu \phi_1(x)$. For the equal-time commutators we
apply the canonical rules, which are satisfied, up to a renormalization constant, 
by the interpolating fields too \cite{Bart}. Then the only nonzero term in
(\ref{degen}) is given by the commutator
\begin{equation}\label{comut}
\delta(x_0) [\phi_1(x),{\mathcal H}_w(0)]\,=
\delta(x_0) [\phi_1(x), \partial_0 \phi_1(0)]
{\delta {\mathcal H}_w\over \delta  \partial_0 \phi_1}\approx \delta (x)\,
j^{(2)}_0\,,
\end{equation}
where  $j^{(2)}_\mu$ is the second current in the weak hamiltonian (for some
terms of ${\mathcal H}_w(0)$ the argument applies after a Fierz rearrangement).
By including this expression in (\ref{degen}), identifying the normalization
constant with the meson decay constant $f_{P_1}$ and restoring the Lorenz
covariance, we notice that the additional term  in the
dispersion relation can be written in the form
\begin{equation}\label{degen1}
 i f_{P_1} k_{1\mu}  \langle P_2(k_2)|j^{(2)}_\mu|B(p)\rangle\,,
\end{equation}
which is the conventional factorization in terms of a form
factor and a meson decay constant. Of course, nonfactorizable corrections
might arise if one goes beyond the simple relation, used in deriving
(\ref{degen1}), between one of the currents in the effective weak  hamiltonian
and the interpolating field of a final meson.

By combining the new term (\ref{degen1}) with the 
dispersive integral obtained from (\ref{lsz1}), we propose the
following dispersion representation for the weak amplitude:
\begin{equation}\label{direl} A(m_B^2, m_1^2, m_2^2)= A^{(0)}(m_B^2, m_1^2,
m_2^2)\,+\, {1\over \pi} \int\limits_0^{(m_B-m_2)^2}{\rm d}z {\mbox{Disc}\,
A(m_B^2, z, m_2^2)\over z-m_1^2-i\epsilon}\,. \end{equation} As discussed
above, the first term  can be evaluated approximately 
using conventional factorization, while in the second one the
dispersion variable is the mass of the meson which does not contain the
spectator quark. The representation (\ref{direl}) gives, when the final
state interactions are switched-off, a term  which can be approximated by
the factorized amplitude (with possible hard scattering corrections), which
is a reasonable consistency condition \footnote {We thank M. Beneke for
emphasizing this point.}. We are aware that we have not given a rigorous proof
for the dispersion relation (\ref{direl}), but brought some general arguments,
supported by the consistency of the physical picture.

Below, we shall investigate in more detail the evaluation of the
above dispersion relation. A first remark is that in
the unitarity sum (\ref{spectr}) one can use as a complete set of hadronic
states $|n\rangle$ either the ``in'' or the ``out'' states.  The equivalence of
these two sets was used in \cite{GoTr} to prove the reality of the spectral
function for $T$ (or $CP$) conserving interactions.
Let us consider what happens if the weak hamiltonian contains
a  $CP$ violating part. In the standard model the weak hamiltonian
${\mathcal H}_w$ has the general form
\begin{equation}\label{hweak}
{\mathcal H}_w={\cal O}_1+{\cal O}_2\, {\rm e}^{ i \gamma} + h.c. \,,
\end{equation}
where $ {\cal O}_j$,  $j=1,2$, are  products of vector and axial weak currents
involving only real coefficients, and $\gamma$ is the weak angle
of the Cabibbo-Kobayashi-Maskawa (CKM) matrix in the standard parametrization
($\gamma =\mbox{Arg}(V^*_{ub}))$. Then the spectral function defined in
Eq.~(\ref{spectr}) can be written as
\begin{equation}\label{sigma12}
\mbox{Disc}\, A(m_B^2, z, m_2^2)\,=\sigma_1 (z)+\sigma_2 (z)\,
{\mathrm e}^{i \gamma} \,,
\end{equation}
where $\sigma_1 (z)$ ($\sigma_2 (z)$) are obtained by replacing  ${\cal H}_w$
in the r.h.s. of Eq.~(\ref{spectr}) with  ${\cal O}_1$ (${\cal O}_2$),
respectively (we took for convenience a process involving the weak phase
$\gamma$). It is easy to show that  $\sigma_1 (z)$ and $\sigma_2 (z)$  are real
functions. Indeed, let  us assume that a complete set  $|n, {\mathrm in}\rangle$
is inserted in the unitarity sum of Eq.~(\ref{spectr}).
Following  \cite{GoTr},  we can express the two matrix elements in this sum as
\begin{equation}\label{etapt}
\langle P_2(k_2) |\eta_1| n, {\mathrm in}\rangle =
\langle P_2(k_2)|(PT)^{-1}(PT) \eta_1 (PT)^{-1}(PT)|n, {\mathrm in}\rangle =
\langle P_2(k_2)|\eta_1|n, {\mathrm out}\rangle ^*\,
\end{equation}
and
\begin{equation}\label{hweakpt}
\langle n, {\mathrm in} |
{\cal O}_j |  B(p)\rangle=
\langle n, {\mathrm in}|(PT)^{{-1}}(PT) {\cal O}_j (PT)^{{-1}} (PT)|
B(p)\rangle
\label{hpt1}\, =\! \langle n, {\mathrm out}|{\cal O}_j| B(p)\rangle^*\,.
\end{equation}
We used here the transformation properties of the  ${\cal O}_j$
operators  under $P$ and $T$ transformations, and the fact
that under space-time reversal the particles conserve their momenta,
the in (out) states becoming out (in) states, respectively. Moreover,
the matrix elements are replaced by their complex conjugates,
given the antiunitary character of the operator $T$.
By using the relations (\ref{etapt}) and (\ref{hweakpt}) in
Eq.~(\ref{spectr}) we obtain
\begin{eqnarray}\label{reality}
\sigma_j (z) &=& {1\over 2\sqrt{2\omega_1}}\sum_{n}\delta  (k_1+k_2-p_n)
\langle
P_2(k_2) |
\eta_1|n, {\mathrm in}\rangle\langle n, {\mathrm in}|
{\cal O}_j|  B(p)\rangle \,\nonumber\\
&=&{1\over 2 \sqrt{2\omega_1}}\left [\sum_{n}\delta  (k_1+k_2-p_n)  \langle
P_2(k_2) |
\eta_1|n, {\mathrm out}\rangle\langle n, {\mathrm out}|
{\cal O}_j|B(p)\rangle \right ]^*\nonumber \\
&=&\sigma_j^*(z)\,,\,\,j=1,2\, ,
\end{eqnarray}
where the equivalence between the complete sets of in and out states
in the definition of $\sigma (z)$ is taken into account.
The Eqs. (\ref{sigma12}) and (\ref {reality}) express in a detailed form
the reality properties of the discontinuity, and bring a
correction to the relation (16) given in Ref. \cite{CaMi}.

{}From Eq. (\ref{reality}) it follows
that the discontinuities $\sigma_j(z)$ are manifestly real only if the
intermediate states form  a complete set. If the unitarity
sum is truncated, this property is lost, since various terms have
complex phases which do not compensate each other in an obvious way. By
inserting in Eq. (\ref{spectr}) a set of states  $|n,\,{\mathrm out}\rangle$,
we obtain for each term the product of the weak amplitudes with the complex
conjugates of the strong amplitudes. If the set which is
inserted consists of  $|n,{\mathrm in}\rangle$ states, then in Eq.
(\ref{spectr}) the strong amplitudes appear as such, while in  the weak
amplitudes we must take the complex conjugate of the strong phases (the weak
phase multiplying the operator ${\cal O}_2$ in (\ref{hweak}) is unmodified,
since the same part of the weak hamiltonian acts on both in and out states).

As noticed in Ref. \cite{GoTr},  it is convenient to write  the
complete set of states  $|n\rangle$ as a combination
$1/2|n, {\mathrm in}\rangle +1/2|n, {\mathrm out}\rangle$. In the case of $CP$
conserving interactions this procedure maintains the reality
condition of the spectral function  (this method was used in the so-called
Omn\`es solution for the electromagnetic form factor \cite{Omnes}). In our case
it is easy to show that the Goldberger-Treiman procedure respects at
all stages of approximation  the specific reality conditions expressed in the
relations  (\ref{sigma12}) and (\ref{reality}).

We consider now the two-particle approximation, when the
dispersion relation takes a very simple form. Indeed, in this case
the on shell weak decay amplitudes $A_{B\to P_3 P_4}$ appearing in the
unitarity sum (\ref{spectr}) are independent of the phase space integration
variables, and also of the dispersion variable $z$. Therefore, the dispersion
representation (\ref{direl}) becomes an algebraic relation among on shell weak
amplitudes \cite{CaMi}.

If we insert in Eq. (\ref{spectr}) a set of states  
$|P_3P_4, {\mathrm out}\rangle$, we obtain
\begin{equation}\label{sumout}
\mbox{Disc}\, A_{B\to P_1P_2}
=\sum _{\{P_3P_4\}} C^*_{P_3P_4;P_1P_2}(z)
A_{B\to P_3 P_4}\,,
\end{equation}
where $ C^*_{P_3P_4;P_1P_2}(z)$ are the complex conjugates of the coefficients
\begin{equation}\label{coef}
 C_{P_3P_4;P_1P_2}(z)
={1\over 2}\int
{{\mathrm d}^3{\mathbf k}_3\over(2\pi)^{3} 2\omega_3}{{\mathrm d}^3{\mathbf
k}_4\over(2\pi)^{3}2\omega_4}
(2\pi)^4\delta^{(4)}(p-k_3-k_4) {\mathcal M}_{P_3 P_4 \to P_1 P_2}(s,t)\,,
\end{equation}
defined in terms of the strong amplitudes $ {\mathcal M}_{P_3 P_4 \to P_1
P_2}(s,t)$, where $s$, $t$ and $u$ are the Mandelstam variables.
These coefficients depend on the masses of all the particles participating
in the rescattering process, in particular, they depend on the dispersive
variable  $z = k^2_1$.

Similarly, by including in Eq. (\ref{spectr})  a set of states 
$|P_3P_4,\, {\mathrm in}\rangle$ we obtain
\begin{equation}\label{sumin}
\mbox{Disc}\, A_{B\to P_1P_2}
=\sum _{\{P_3P_4\}} C_{P_3P_4;P_1P_2}(z)
\bar A_{B\to P_3 P_4}\,,
\end{equation}
where, according to the above discussion, the amplitude $\bar A_{B\to P_3 P_4}$
is obtained from  $A_{B\to P_3 P_4}$ by changing the sign of the strong phase,
namely,
$$\bar A_{B\to P_3 P_4}=\vert A_{B\to P_3 P_4}\vert\, {\mathrm e}^{-i \phi_s}\,
{\mathrm e}^{i \phi_w}\,, $$
where  $\phi_s$ ($\phi_w$) denotes the strong (weak) phase.

Now by performing the  symmetric  Goldberger-Treiman summation as explained
above,  we obtain, instead of (\ref{sumout}) or (\ref{sumin}), the expression
\begin{equation}\label{sum1}
\mbox{Disc}\,  A_{B\to P_1P_2}  ={1\over 2}\sum _{\{P_3P_4\}}
C_{P_3P_4;P_1P_2}(z)\bar A_{B\to  P_3 P_4}+ {1\over 2} \sum_{\{P_3P_4\}}
C^*_{P_3P_4;P_1P_2}(z) A_{B\to P_3 P_4}\,.  \end{equation}
With this discontinuity, the dispersion relation (\ref{direl}) becomes
\begin{equation}\label{sumint1}
 A_{B\to P_1P_2}
= A^{(0)}_{B\to P_1P_2}\,+\,{1\over 2}\sum _{\{P_3P_4\}} \Gamma_{P_3P_4;P_1P_2}
\bar A_{B\to P_3 P_4}+ {1\over 2}\sum_{\{P_3P_4\}}
\overline\Gamma_{P_3P_4;P_1P_2}
A_{B\to P_3 P_4}\,,
\end{equation}
where $A^{(0)}_{B\to P_1P_2}$ can be approximated by the amplitude in the
factorization limit, and the coefficients  $\Gamma_{P_3P_4;P_1P_2}$ and
$\overline\Gamma_{P_3P_4;P_1P_2}$
are defined as
\begin{equation}\label{gama}
\Gamma_{P_3 P_4; P_1 P_2}={1\over \pi}
\int\limits_0^{(m_B-m_2)^2}{\mathrm d}z{C_{ P_3P_4;
P_1P_2}(z)\over z-m_1^2-i
\epsilon}\,,
\end{equation} and
\begin{equation}\label{bargama}
\overline\Gamma_{P_3 P_4;P_1 P_2}={1\over \pi}
\int\limits_0^{(m_B-m_2)^2}{\mathrm d}z{C^*_{ P_3P_4;
P_1P_2}(z)\over z-m_1^2-i
\epsilon}\,.
\end{equation}
The Eqs. (\ref{sum1})-(\ref{bargama}) are the result of the Goldberger-Treiman
procedure in the presence of $CP$ violating interactions, replacing
the corresponding  relations (38) and (39) given in \cite{CaMi}.

The strong amplitudes ${\mathcal M}_{P_3 P_4;P_1 P_2}(s,t)$ entering the
expression  (\ref{coef}) of the coefficients $C_{ P_3P_4;P_1P_2}(z)$ are
evaluated at the c.m. energy squared $s=m_B^2$, which, as emphasized in
\cite{Dono1}, is high enough to justify the
application of  Regge theory \cite{Coll}. The amplitudes 
${\mathcal M}_{P_3 P_4;P_1 P_2}(s,t)$  
can be expressed  therefore as sums over Regge exchanges in
the crossed channels, more exactly \cite{Dele}: near the forward direction
(small $t$) the $t$-channel exchanges are taken into account, while near the
backward direction (small $u$) the $u$-channel exchanges are considered. The
standard form of a Regge amplitude given by a trajectory exchanged in the
$t$-channel is
\begin{equation}\label{regge0}
-\gamma (t)\,{\tau + {\mathrm e}^{-i\pi\alpha(t)}\over \sin
\pi\alpha(t)}\,\left({s\over s_o}\right )^{\alpha(t)}\,,
\end{equation}
where $\gamma (t)$ is the residue function, $\tau$ the signature, 
$\alpha(t) =\alpha_0 +\alpha' t$ the linear trajectory, and 
$s_0\approx 1\, {\mathrm GeV}^2$. 
A Regge trajectory exchanged in the $u$-channel gives an expression
similar to (\ref{regge0}), with $t$ replaced by $u$.

Using the signatures $\tau=1$ for $C=1$ trajectories and  $\tau=-1$ for $C=-1$,
we express the strong amplitude  near the foward direction as \cite{CaMi}
\begin{eqnarray}\label{regget}
{\mathcal M}_{P_3 P_4;P_1 P_2}(s,t)=-\sum_{V=P, f, A_2, K^{**}...}
\gamma^V_{P_3 P_4;P_1 P_2} (t){{\mathrm
e}^{-i{\pi\alpha_V(t)\over 2}}\over \sin{\pi\alpha_V (t)\over 2}}
\left({s\over s_0}\right)^{\alpha_V(t)}\,\nonumber\\
+\sum_{V=\rho, K^*...}i
\gamma ^V_{P_3 P_4\to P_1 P_2}(t){{\mathrm e}^{-i{\pi
\alpha_V(t)\over 2}}\over \cos{\pi\alpha_V(t)\over 2}}\left({s\over s_0}
\right)^{\alpha_V(t)}\,,
\end{eqnarray}
where the sum extends over the $t$-channel poles. The first sum includes the
Pomeron (which contributes only to the elastic scattering) and tensor
particles, the second sum includes vector particles.
In Ref. \cite{CaMi} we used this expression in the integral over the phase
space in Eq. (\ref{coef}), which finally reduces to an integral over the c.m.
scattering angle $\theta$. Since the Regge amplitudes (\ref {regget}) decrease
exponentially at large $t$,  the dominant contribution in the integral  is
brought by the forward region. This gives a correct result for amplitudes which
are small near the backward direction (for exemple, in processes where the $u$
channel is exotic), otherwise, it misses in general the contribution due to
large angles.
As discussed in \cite{Dele}, it is more appropriate to separate the integration
over the scattering angle  $\theta$ in two regions, one for small angles using
the Regge expression (\ref{regget}), the other for large angles where a similar
expression
\begin{eqnarray}\label{reggeu}
{\mathcal M}_{P_3 P_4;P_1 P_2}(s,t)=-\sum_{V= f, A_2, K^{**}...}
\gamma^V_{P_3 P_4;P_1 P_2} (u){{\mathrm
e}^{-i{\pi\alpha_V(u)\over 2}}\over \sin{\pi\alpha_V (u)\over 2}}
\left({s\over s_0}\right)^{\alpha_V(u)}\,\nonumber\\
+\sum_{V=\rho, K^*...}i
\gamma ^V_{P_3 P_4\to P_1 P_2}(u){{\mathrm e}^{-i{\pi
\alpha_V(u)\over 2}}\over \cos{\pi\alpha_V(u)\over 2}}\left({s\over s_0}
\right)^{\alpha_V(u)}\,,
\end{eqnarray}
given by the $u$-channel Regge trajectories is valid. Following \cite{Dele}, in
performing the unitarity integral (\ref{coef}) we  adopt the expression
(\ref{regget}) of the amplitude ${\mathcal M}_{P_3 P_4;P_1 P_2}(s,t)$ for
$\cos\theta >0$ and the expression (\ref{reggeu})  for
$\cos\theta <0$. Assuming  the residua functions
$\gamma^V_{P_3 P_4;P_1 P_2} (t)$ and  $\gamma^V_{P_3 P_4;P_1 P_2} (u)$ to be
constant along the relevant integration ranges, and neglecting also the $t
(u)$ dependence of the denominators in (\ref{regget}) and (\ref{reggeu}),
the integration over the momenta
${\mathbf k}_3$ and ${\mathbf k}_4$ in Eq. (\ref{coef}) is straightforward,
knowing the kinematic relations between the  Mandelstam variables $t$ and $u$
and the scattering angle.
The coefficients $C_{ P_3P_4; P_1P_2}$ can be expressed as
\begin{equation}\label{coef1}
C_{ P_3P_4; P_1P_2}(z)= \sum_{\{V_t\}}\xi_{V_t}
\gamma^{V_t}_{P_3 P_4;P_1 P_2} \, \kappa^{V_t}_{P_3 P_4;P_1 P_2}(z)+
\sum_{\{V_u\}}\xi_{V_u}
\gamma^{V_u}_{P_3 P_4;P_1 P_2} \, \kappa^{V_u}_{P_3 P_4;P_1 P_2}(z) \,,
\end{equation}
where the first (second) sum includes the contribution of the $t$($u$)-channel
trajectories. In Eq. (\ref{coef1}), $\xi_V$ is a numerical factor due to the
signature (equal to: $-1$ for the Pomeron, $i\sqrt{2}$ for $C=-1$ trajectories,
and $-\sqrt{2}$ for $C=1$ physical trajectories). The coefficients
$\kappa^{V_t}_{P_3 P_4;P_1 P_2}(z)$ appearing in the first sum have the
expression
\begin{equation}\label{kappa}
\kappa^{V_t}_{P_3 P_4;P_1 P_2}(z)= { k_{34} \over 16 \pi m_B}
{\mathcal R}_{V_t}^{-1}(z)
\left[{\mathrm e}^{{\mathcal R}_{V_t}(z)}-1\right]
\exp\left[\left(\alpha_{0,{V_t}}+\alpha'_{V_t}t_0(z)\right)
\left(\ln {m_B^2\over s_0}-i{\pi\over2}\right)\right]\,,
\end{equation}
obtained by integration over the region $0<\cos\theta<1$ of the phase space.
We used the notation \cite{CaMi}
\begin{equation}\label{R}
{\mathcal R}_V(z)= 2\alpha'_Vk_{12}(z)
k_{34}\left(\ln{m_B^2\over s_0}-i{\pi\over2}\right)\,,
\end{equation}
where $k_{12}$ and  $k_{34}$ denote the c.m. three momenta and
\begin{equation}\label{t0}
t_0(z)=z+m_3^2-{(m_B^2+m_3^2-m_4^2)(m_B^2+z-m_2^2)\over2m_B^2}\,.
\end{equation}
As for the coefficients $\kappa^{V_u}_{P_3 P_4;P_1 P_2}(z)$ appearing
in the second sum of Eq. (\ref{coef1}), they are
given by the region $-1<\cos\theta<0$ of the phase space integral, and their
expression is similar to (\ref{kappa}), with the $t$-channel trajectories
replaced by the $u$-channel trajectories, and $t_0(z)$ replaced
by the variable $u_0(z)$, which is obtained from (\ref{t0}) by interchanging
the places of $m_3$ and $m_4$.
The new relations (\ref{coef1})-(\ref{kappa}) improve the corresponding
expressions Eqs. (28)-(30) given in Ref. \cite{CaMi}.

\section{Applications to the  $B^0\to\pi^+\pi^-$ and $B^+\to \pi^0K^+$  decays}

An application of the dispersive formalism to the decay $B^0\to \pi^+\pi^-$ was
already  discussed in Ref. \cite{CaMi}. In the present work, we reconsider this
analysis using the improvements presented above.
As intermediate states $\{P_3P_4\}$ in  the dispersion relation (\ref{sumint1})
we include $\pi^+\pi^-$ for the elastic rescattering, and two-particle states
responsible for the soft inelastic rescattering. We take into account  the
contribution of the lowest pseudoscalar mesons:
$\pi^0\pi^0$, $ K^+ K^-$, $ K^0\bar K^0$,
$\eta_8 \eta_8$,  $\eta_1 \eta_1$ and  $\eta_1 \eta_8$. Here $\eta_8$ and
$\eta_1$ denote the $SU(3)$  octet and singlet, respectively, and we assumed
for simplicity that the mixing is negligible (we mention that
the singlet $\eta_1$  was not included in the previous analysis \cite{CaMi}).
Then, assuming $SU(3)$ flavor symmetry, all the $B$ decay amplitudes entering
the dispersion relation can be  expressed in terms of a certain set of
amplitudes associated to quark diagrams \cite{Gron}.  Following \cite{Gron},
we shall assume that the annihilation, penguin annihilation, electroweak
penguin and exchange diagrams are negligible. For
simplicity, we also neglect in this first step the tree color suppressed
amplitude, as in Ref. \cite{CaMi}. Of course, the final state
interactions can modify the naive estimates based on quark diagrams 
\cite{BlokG}, therefore, the analysis presented below is only a first 
approximation to be refined in a future work. With these assumptions 
$A_{B^0\to K^+K^-}=0$, and the remaining amplitudes entering the
dispersion relation have the expressions
\begin{eqnarray}\label{notation}
&&A_{B^0\to \pi^+\pi^-}=
-(A_T\,{\mathrm e}^{i\gamma}+A_P\,{\mathrm e}^{-i\beta})\,,\quad
A_{B^0\to \pi^0\pi^0}={1\over\sqrt{2}}
A_P\,{\mathrm e}^{-i\beta}\,,\nonumber\\
&&A_{B^0\to K^0\bar K^0}=A_P~{\mathrm
e}^{-i\beta}\,,\quad
A_{B^0\to\eta_{8}\eta_{8}}={1\over3\sqrt{2}} A_P~{\mathrm e}^{-i\beta}
\,,\nonumber\\
&&A_{B^0\to\eta_{1}\eta_{1}}={\sqrt{2}\over3}
A_P~{\mathrm e}^{-i\beta} \,,\quad
A_{B^0\to\eta_{8}\eta_{1}}=-{\sqrt{2}\over3}A_P~{\mathrm e}^{-i\beta} \,,
\end{eqnarray}
the weak phases  being defined as $\beta= {\mathrm Arg}(V^*_{td})$ and
$\gamma= {\mathrm Arg}(V^*_{ub})$.

The determination of
the Regge residua $\gamma^V_{P_3 P_4;P_1 P_2}$ which appear in the expression
(\ref{coef1}) of
the coefficients $C_{P_3P_4;P_1;P_2}$  was described in detail in \cite{CaMi}.
Using the optical theorem and the usual Regge parametrization of the total
hadronic cross sections we obtain the following values
\begin{equation}\label{resid}
\gamma^P_{\pi^+\pi^-;\pi^+\pi^-} \equiv \gamma_P^2= 25.6\,,\quad
\gamma^{\rho}_{\pi^+\pi^-;\pi^+\pi^-}\equiv\gamma_1^2=31.4\,,\quad
\gamma^{f_8}_{\pi^+\pi^-;\pi^+\pi^-}\equiv\gamma_2^2=35.3\,,
\end{equation}
for the residua of the Pomeron, the $\rho$ and the $f$ mesons, respectively.
All the other residua are expressed in terms of the couplings $\gamma_1^2$ and
$\gamma_2^2$ by flavour $SU(3)$ symmetry. For completeness we give their values
in Table 1, where we indicate for each trajectory: the result given by the
Clebsch-Gordan coefficients,  the sign due to the relation
between the quark content of the mesons and the $SU(3)$ vector assigned
to them (with the convention of Ref.~\cite{Gron}), and the Regge factor $\xi_V$
defined below Eq. (\ref{coef1}). The quantity $\xi_V\,
\gamma^V_{P_3P_4;P_1P_2}$ appearing  in (\ref{coef1}) is obtained for each
trajectory by taking the product of the values in the last three columns of
Table 1.

As follows from Table 1, the process $\pi^0\pi^0\to\pi^+\pi^-$  does not
contribute actually to the dispersion relation, since the contribution of the
$\rho$ trajectory in the $t$-channel is exactly compensated in the unitarity
integral by an equal term given by the $u$-channel.
On the contrary, in the case of the  $A_2$ trajectory,
the contributions in the $t$ and $u$ channels are equal and add to
each other.  We also note that the couplings of the singlet $\eta_1$ given in
Table 1 are obtained by assuming an  exact $U(3)$ symmetry. We took into
account in the numerical calculations that deviations 
from these values are possible due to the $U(1)$ anomaly.

With the input described above one can calculate easily the coefficients
(\ref{coef1}) and the dispersive integrals (\ref{gama})-(\ref{bargama})
giving the coefficients $\Gamma$ and $\overline\Gamma$. By inserting these
coefficients and the explicit expressions (\ref{notation}) of the decay
amplitudes in the dispersion relation  (\ref{sumint1}), we derive an
algebraic equation involving the complex quantities $A_T$ and $A_P$. Let us
denote $R = \vert A_P/ A_T \vert$ and $\delta=\delta_P-\delta_T$, where
$\delta_T$ ($\delta_P$) is the strong phase of  $A_T$ ($A_P$), respectively.
Then, after dividing  both sides of the relation (\ref{sumint1}) by $-|A_T|$
and multiplying  by $ {\mathrm e}^{-i\delta_T}$, we obtain the following
equation:
\begin{eqnarray}\label{drpi}
& &{\mathrm e}^{i\gamma}+ R {\mathrm e}^{i\delta}{\mathrm e}^{-i\beta} =
{{\mathrm e}^{-i\delta_T}\over A_T} [A_T^{(0)} {\mathrm e}^{i\gamma}+A_P^{(0)}
{\mathrm e}^{-i\beta}]\nonumber\\
& &-\left[(0.01 +1.27\,i) +(0.75 - 1.01 \,i)\, {\mathrm
e}^{-2i\delta_T}\right]\, {\mathrm e}^{i\gamma}\nonumber\\
& &+R\left[ -(1.97 +2.64\,i)\, {\mathrm e}^{i\delta}{\mathrm e}^{-i\beta}
-(1.78-1.99 i)\, {\mathrm e}^{-i\delta}{\mathrm e}^{-i\beta} 
{\mathrm e}^{-2i\delta_T}\right]\,.
\end{eqnarray}
where $A_T^{(0)}$ and  $A_P^{(0)}$ are the tree and penguin amplitudes in the
factorization approximation.
We mention that in Ref. \cite{CaMi} the equation similar to (\ref{drpi}) did
not contain the factorized amplitude and, due to an improper application
of the Goldberger-Treiman technique, the weak phase $\beta$  appearing 
in the last term had a wrong sign.

Multiplying both sides of Eq. (\ref{drpi}) by ${\mathrm e}^{i\beta}$ one
notices that the weak angles appear in the combination
$\gamma+\beta=\pi-\alpha$, where $\alpha$ is the third angle of the unitarity
triangle. Then, solving  the complex equation  for  $R$ and  $\alpha$  we
derive the expressions
\begin{equation}\label{Rpi}
R(\delta_T, \delta)= \left\vert{1.010+1.27\,i +(0.75 - 1.01 \,i)
{\mathrm e}^{-2i\delta_T}-{{\mathrm e}^{-i\delta_T}\over A_T}\left[A_T^{(0)}
+A_P^{(0)} {\mathrm e}^{-i(\gamma+\beta)}\right]\over -(1.97 +2.64\,i)\,
{\mathrm e}^{i\delta}-(1.78 -1.99 i)\, {\mathrm e}^{-i\delta} 
{\mathrm e}^{-2i\delta_T}}\right\vert\,,
\end{equation} and
\begin{eqnarray}\label{alpha}
\alpha(\delta_T, \delta)=\pi+ {\mathrm Arg}\left[1.01 +1.27\,i 
+(0.75 - 1.01 \,i)\,
{\mathrm e}^{-2i\delta_T}-{A_T^{(0)}\over A_T}
{\mathrm e}^{-i\delta_T}\right]\nonumber \\-{\mathrm Arg}
\left[-(1.97+2.64\,i)\,
{\mathrm e}^{i\delta}-(1.78 -1.99 i)\, {\mathrm e}^{-i\delta} 
{\mathrm e}^{-2i\delta_T}+{{\mathrm e}^{-i\delta_T}\over R} {A_P^{(0)}\over A_T}
\right]\,,
\end{eqnarray}
where in the last equation we use $R$ from Eq. (\ref{Rpi}). 
The evaluation of these expressions requires
the knowledge of the ratios $A_P^{(0)}/A_T^{(0)}$ and $A_T^{(0)}/A_T$. We use
for illustration $A_P^{(0)}/A_T^{(0)}=0.08$ \cite{BeBu}, and a reasonable
choice $A_T^{(0)}/A_T\approx 0.9$. 
The expression of $R$ contains also the weak angle
$\beta+\gamma$, but the dependence on this parameter is very weak. In Fig. 1
we represent  $R$ (Eq. (\ref{Rpi})) as a function of the phase
difference $\delta$, for two values of $\delta_T$. We recall that the
ratio $R$ is expected to be less than one,
and such values are obtained for both $\delta_T=0$ and $\delta_T=\pi/12$.
The ratio $R$ is actually a periodic function of $\delta$ with a period equal
to $\pi$, which implies that discrete ambiguities
affect the determination of this phase difference for a given value of $R$.

In Fig. 2 we show  $\alpha$ as a function of
$\delta$, for  $\delta_T=0$ and $\delta_T=\pi/12$ .
Since one expects positive values of $\alpha$, the curves shown in Fig. 2
indicate that negative  values of  $\delta$ are preferred. We have checked
that the results are rather stable with respect to the variation of the
$\eta_1$ couplings. For instance, by varying  the corresponding Regge residua
from 0 up to a value 2 or 3 times larger than the value given in  Table 1, we
notice that the curves in Figs. 1 and 2 are slightly shifted, but the general
behavior remains the same. Actually, the dominant contribution is given by the
elastic channel, more precisely by the Pomeron, as is seen in Fig. 1,
where we show the ratio $R$ for  $\delta_T=\pi/12$, keeping only the
contribution of the Pomeron in the Regge amplitudes (dotted curve).

For comparison, in Ref. \cite{CaMi} the corresponding equation did not contain
the amplitude in the factorization limit, and instead of
the angles $\beta+\gamma$ and $\delta$, we had the angles $\gamma$ and
$\delta-\beta$, respectively (see the remark below Eq. (\ref{drpi})).
The Figs. 1 and 2 given in \cite{CaMi} become therefore  meaningful if
the arguments are replaced accordingly; they represent in fact $R$ and $\delta$
as functions of $\beta+\gamma=\pi-\alpha$, neglecting the factorized term. The
values of $R$ obtained for  $\delta_T=0$ were larger than one, suggesting that
small values of  $\delta_T$ are excluded. Now, as seen in Fig.1, 
the improved dispersion representation proposed in the present work 
indicates that reasonable values of $R$ are compatible 
with small values of $\delta_T$.

In a second application we consider the decay  $B^+\to \pi^0K^+$, taking as
intermediate states $P_3P_4$ the pseudoscalar mesons $ \pi^0K^+$,  $\pi^+K^0$,
$\eta_8 K^+$ and $\eta_1K^+$, allowed by the strong interactions. A model
independent analysis  based on isospin symmetry done in Ref.~\cite{Neub}
gives for the amplitude of $B^+\to \pi^0K^+$ decay the expression
\begin{equation}\label{a1}
A_{B^+\to \pi^0K^+}=-{P\over \sqrt{2}}\left[1-\epsilon_a e^{i\gamma}\,
e^{i\eta}-\epsilon_{3/2} e^{i\phi} (e^{i\gamma}-\delta_{EW})\right]\,,
\end{equation}
where $P$ denotes the dominant penguin amplitude, $\delta_{EW}$ is an
electroweak correction and $\epsilon_a, \epsilon_{3/2}, \eta$  and $\phi$ are
hadronic parameters ($\phi = \phi_{3/2}-\phi_P$, where $ \phi_{3/2}$ is the
strong phase of the $I=3/2$ amplitude and $\phi_P$ the phase of $P$) .
According to \cite{Neub}, the term  proportional to $\epsilon_a$, which is due
to non dominant penguin and annihilation topologies, is smaller than the last
terms appearing in Eq. (\ref{a1}).  Neglecting in a first approximation this
term and using flavour  $SU(3)$ symmetry for the weak decays \cite{Gron}, we
write the amplitudes which contribute to the dispersion relation  as
\begin{eqnarray}\label{apik}
&& A_{B^+\to \pi^0K^+}=-{P\over \sqrt{2}}\left[1-r e^{i\phi}\right]\,,\quad
A_{B^+\to \pi^+K^0}=P\,,\nonumber\\
&&A_{B^+\to \eta_8 K^+}= {P\over \sqrt{6}}\left[1+r e^{i\phi}\right]\,,\quad
A_{B^+\to \eta_1 K^+}=- {P\over \sqrt{3}}\left[2- r e^{i\phi}\right]\,,
\end{eqnarray} where
\begin{equation}\label{rpik}
r=\epsilon_{3/2} \left({\mathrm e}^{i\gamma}-\delta_{EW}\right)\,.
\end{equation}
We need also the amplitude $ A^{(0)}_{B^+\to \pi^0K^+}$ in the factorized
approximation, which we write as \cite{Falk}
$$ A^{(0)}_{B^+\to \pi^0K^+}\approx-{P^{(0)}\over \sqrt{2}}\left[1-0.35
e^{i\gamma} \right]\,.$$
The Regge residua entering the coefficients (\ref{coef1}) can be expressed in
terms of the same parameters $\gamma_P^2, \gamma_1^2$ and $\gamma_2^2$ 
as defined in (\ref{resid}) by using $SU(3)$. 
We fix the Pomeron coupling to the same value as in the $\pi\pi$
case, and take for the other trajectories the values listed in Table 2, where
the meaning of the columns is the same as in Table 1.

By inserting the amplitudes (\ref {apik}) and the new coefficients $\Gamma$
and $\overline\Gamma$, calculated as above, in the dispersion relation
(\ref{sumint1}), we obtain after simplifying with $|P|$  a complex equation
involving the parameters $r, \phi_P$ and $\phi$. From this equation we obtain
for $r$ the expression
\begin{equation}\label{rpik1}
r(\phi_P, \phi) = (0.12 + 0.09 \,i){
(5.48+ 4.05\, i)- (6.98 - 4.8\, i){\mathrm e}^{2i\phi_P} + i (-3.33 +{\mathrm
e}^{i\gamma}){A_P^{(0)}\over A_P}
{\mathrm e}^{i\phi_P}  \over
(0.16 + 0.75\, i)\, {\mathrm e}^{-i\phi} - {\mathrm e}^{i\phi}\, {\mathrm
e}^{2i\phi_P}}\,.
\end{equation}
The expression of $r$ contains also the ratio $A_P^{(0)}/ A_P$ and the weak
angle  $\gamma$ (actually the dependence of $r$ on $\gamma$ is very weak). 
In Fig. 3 we show for
illustration the real and the imaginary parts of $r$ as functions of $\phi$,
for $\phi_P=\pi/6$, using the estimate $A_P^{(0)}/ A_P\approx 0.7$.
From the definition (\ref{rpik}), calculated with the parameters
$\epsilon_{3/2}=0.24$ and  $\delta_{EW}=0.64$ given in Ref. \cite{Neub},
we can extract the weak angle $\gamma$,  defined as
$\gamma={\mathrm Arg}\,  (r/\epsilon_{3/2}+\delta_{EW})$.
In Fig. 4 we show the variation of  $\gamma$ as a function  of $\phi$
for $\phi_P=\pi/6$.
\section{Conclusions}
In the present work we brought several improvements to
the dispersion formalism  proposed in Ref. \cite{CaMi} for investigating the
hadronic parameters in $B$ nonleptonic decays. The main modification consists
in  the discovery of an additional term in the dispersion representation. The
new dispersion relation is given by  Eq. (\ref{direl}), 
where the first term can be approximated by the factorized amplitude, and the
dispersive variable is the mass of the final meson which does not contain the
spectator quark. We mention that (\ref{direl}) is not a subtracted dispersion
relation, the origin of the additional term being an equal time commutator in
the LSZ formalism. 
We also treated more carefully the Goldberger-Treiman  procedure to perform the
unitarity sum, and refined the Regge model by the inclusion of both the
$t$ and $u$-channel trajectories. We emphasize that the analytic continuation
in the external mass could be kept under control in the present context,  
since the Regge dynamics is rather universal (as long as the masses are small 
with respect to the energy).

 With the improvements mentioned above we reconsidered the analysis of the
$B^0\to \pi^+\pi^-$ decay previously made, and discussed also the process
$B^+\to \pi^0K^+$.
In the present calculation we restricted the sum over intermediate states to
the lowest pseudoscalar mesons, and invoked  flavour symmetry to reduce the
number of unknown amplitudes.
The credibility of the results relies on the validity of the
assumption that the  higher states do not contribute in a significant way in
the dispesion relation. There are several arguments in favor of this assumption.
First, our results show that the dominant contribution is given by the Pomeron.
 Higher-mass states are suppressed by the phase space
integration appearing in Eq. (\ref{coef}), since in our formalism $s=m_B^2$.
Finally, one can argue that the effect of the higher states is simulated in a
certain sense  by  the Goldberger-Treiman procedure, since it ensures  a
reality property of the discontinuity, which is normally valid when the
unitarity sum is not truncated.

For simplicity, in the present application of the method we neglected the 
amplitudes suggested to be small by the quark diagrams, which introduces 
some model dependence in the final results. A more complete treatment 
including all the amplitudes is possible, by the simultaneous use of 
several dispersion relations for weak amplitudes correlated through 
rescattering effects.

\vskip 0.3cm  {\bf Acknowledgments:} One of the authors (I.C.) is grateful to
Prof. A. de R\'ujula and the CERN Theory Division for hospitality. 
We wish to thank M. Beneke and J. Donoghue for interesting discussions.
This work was partly
realized in the frame of the Cooperation Agreements between IN2P3
and NIPNE-Bucharest, and  between CNRS and the Romanian Academy.
I.C. and L.M. express their thanks
to the Centre de Physique des Particules de Marseille (CPPM) and the Centre de
Physique  Th\'eorique (CPT) of Marseille for hospitality.

\newpage
\mediumtext
\begin{table}
\caption {Values of the Regge residua of the rescattering amplitudes
in $B^0\to\pi^+\pi^-$:  column II indicates the channel,  III the Regge
trajectories,   IV the coupling given by $SU(3)$ Clebsch-Gordan coefficients, V
the additional sign due to the definition of the meson states,  and  VI the
Regge factor $\xi_V$.}
\label{Table 1}
\begin{tabular}{cccccc}
&&&&&\\
I&II&III&IV&V&VI\\
&&&&&\\
\tableline
$\pi^+\pi^-\to\pi^+\pi^-$&&&&&\\
&t&$\rho$&$\gamma_1^2$&$+$&$i\sqrt{2}$\\
&t&$f_8$&$\gamma_2^2$&$+$&$-\sqrt{2}$\\
&u&exotic&&&\\
\tableline
$\pi^0\pi^0\to\pi^+\pi^-$&&&&&\\
&t&$\rho$&$\gamma_1^2$&$-$&$i\sqrt{2}$\\
&u&$\rho$&$-\gamma_1^2$&$-$&$i\sqrt{2}$\\
\tableline
&&&&&\\
$K^0\bar K^0\to\pi^+\pi^-$&&&&&\\
&t&exotic&&&\\
&u&$K^*$&$-1/2\gamma_1^2$&$-$&$i\sqrt{2}$\\
&u&$K^{**}$&$3/2\gamma_2^2$&$-$&$-\sqrt{2}$\\
\tableline
$\eta_8\eta_8\to\pi^+\pi^-$&&&&&\\
&t&$A_2$&$\gamma_2^2$&$-$&$-\sqrt{2}$\\
&u&$A_2$&$\gamma_2^2$&$-$&$-\sqrt{2}$\\
\tableline
$\eta_1\eta_1\to\pi^+\pi^-$&&&&&\\
&t&$A_2$&$5\gamma_2^2$&$-$&$-\sqrt{2}$\\
&u&$A_2$&$5\gamma_2^2$&$-$&$-\sqrt{2}$\\
\tableline
$\eta_8\eta_1\to\pi^+\pi^-$&&&&&\\
&t&$A_2$&$\sqrt{5}\gamma_2^2$&$-$&$-\sqrt{2}$\\
&u&$A_2$&$\sqrt{5}\gamma_2^2$&$-$&$-\sqrt{2}$\\
\end{tabular}
\end{table}
\mediumtext
\begin{table}
\caption {Values of the Regge residua for the rescattering amplitudes in the
$B^+\to \pi^0 K^+$ case. The meaning of the columns is the same as in Table 1.}
\label{Table 2}
\begin{tabular}{cccccc}
&&&&&\\
I&II&III&IV&V&VI\\
&&&&&\\
\tableline
$\pi^0K^+\to\pi^0 K^+$&&&&&\\
&t&$f_8$&$1/2\gamma_2^2$&$-$&$-\sqrt{2}$\\
&u&$K^*$&$1/4\gamma_1^2$&+&$i\sqrt{2}$\\
&u&$K^{**}$&$3/4\gamma_2^2$&+&$-\sqrt{2}$\\
\tableline
$\pi^+K^0\to\pi^0 K^+$&&&&&\\
&t&$\rho$&$-\sqrt{2}/2\gamma_1^2$&+&$i\sqrt{2}$\\
&u&$K^*$&$-\sqrt{2}/4\gamma_1^2$&$-$&$i\sqrt{2}$\\
&u&$K^{**}$&$-3\sqrt{2}/4\gamma_2^2$&$-$&$-\sqrt{2}$\\
\tableline
$\eta_8K^+\to \pi^0K^+$&&&&&\\
&t&$A_2$&$-\sqrt{3}/2\gamma_2^2$&$-$&$-\sqrt{2}$\\
&u&$K^*$&$\sqrt{3}/4\gamma_1^2$&+&$i\sqrt{2}$\\
&u&$K^{**}$&$-\sqrt{3}/4\gamma_2^2$&+&~$-\sqrt{2}$\\
\tableline
$\eta_1K^+\to \pi^0K^+$&&&&\\
&t&$A_2$&$-\sqrt{15}/2\gamma_2^2$&$-$&$-\sqrt{2}$\\
&u&$K^{**}$&$\sqrt{15}/2\gamma_2^2$&+&$-\sqrt{2}$\\
\end{tabular}
\end{table}
\begin{figure}
\epsfxsize=12cm
\centerline{\epsfbox{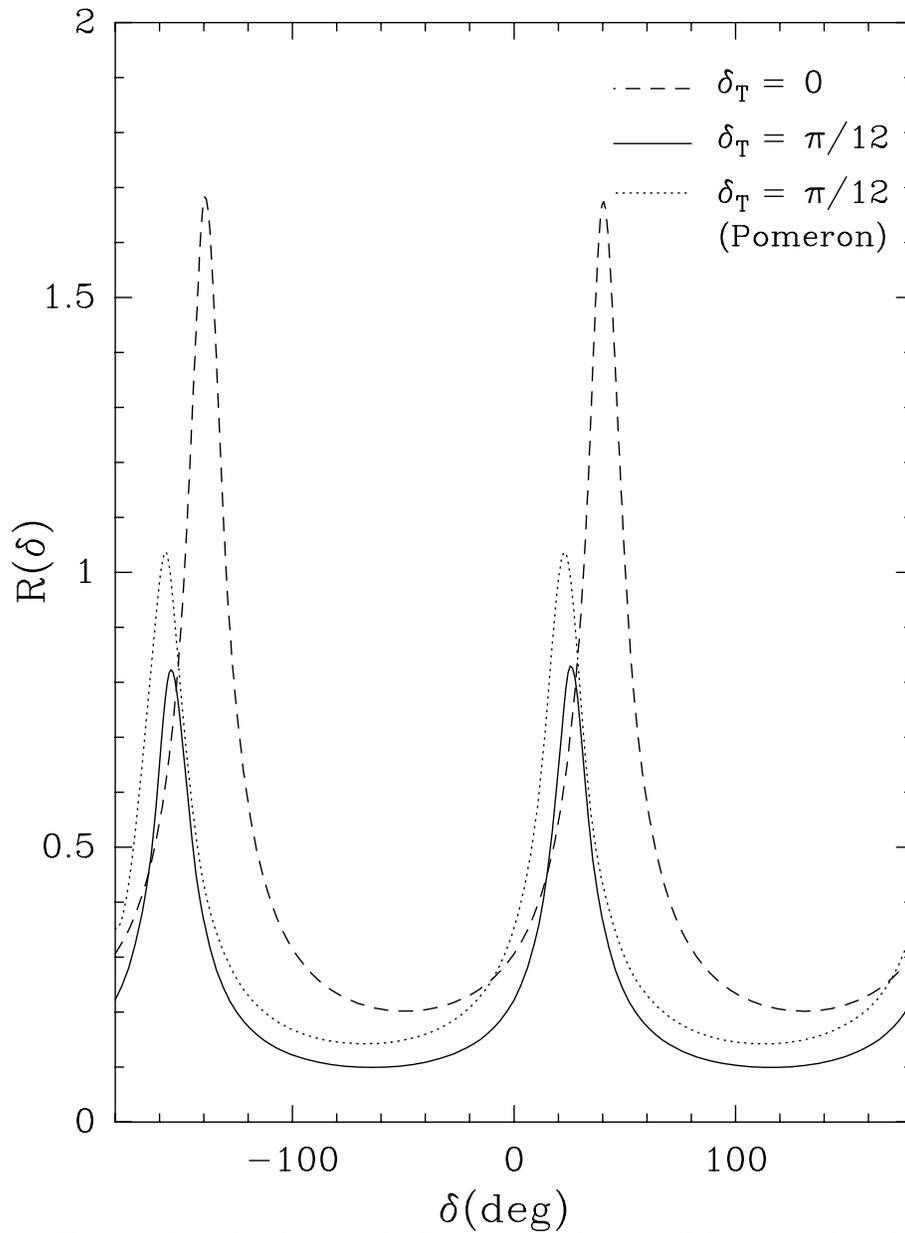}}
\caption{The ratio $R=\vert A_P/ A_T\vert$ given by Eq. (\ref{Rpi}) as a
function of the strong phase difference $\delta$, solid curve
$\delta_T=\pi/12$, dashed curve
$\delta_T=0$. The dotted curve is obtained for $\delta_T=\pi/12$, 
by keeping only the contribution of the Pomeron in the Regge amplitudes.}
\label{fi:rpi}
\end{figure}
\begin{figure}
\epsfxsize=12cm
\centerline{\epsfbox{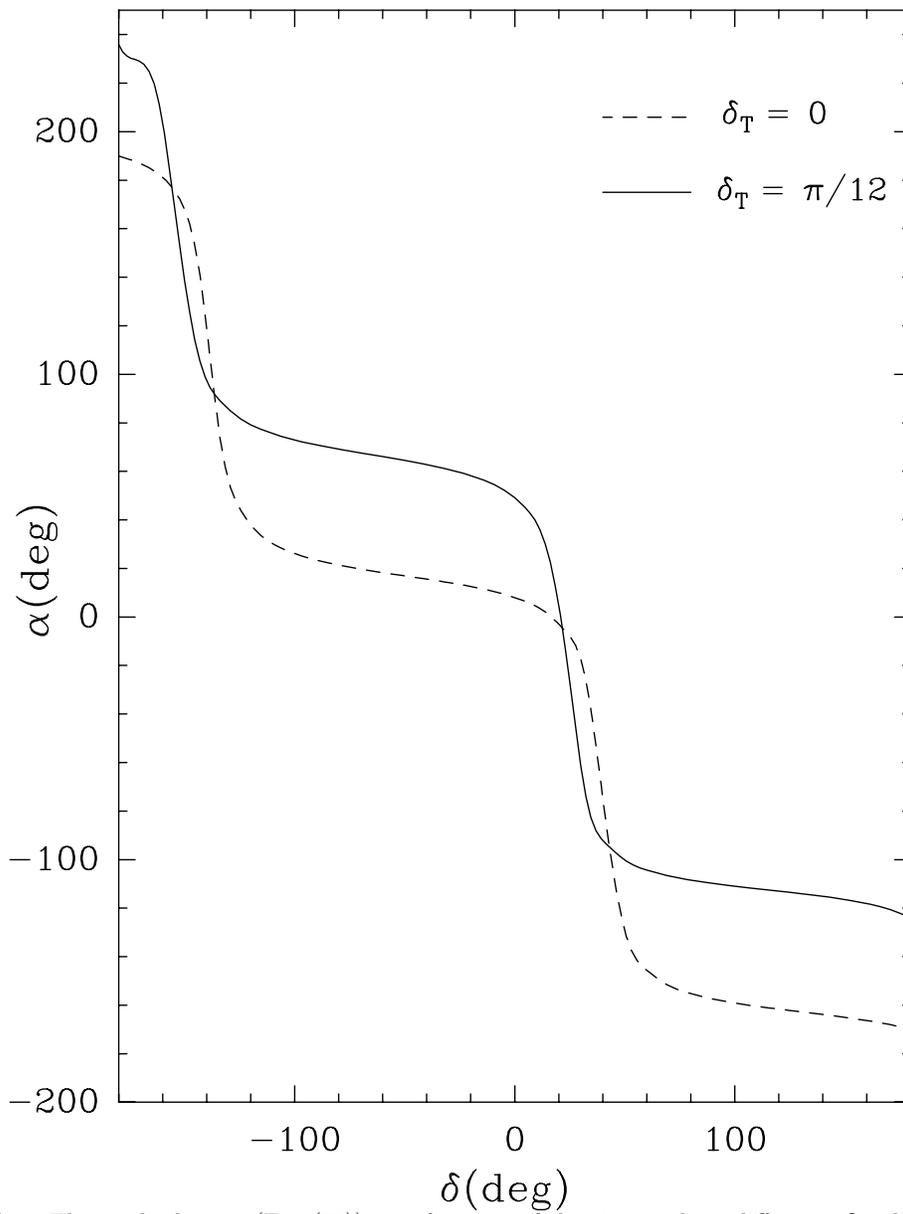}}
\caption{The weak phase $\alpha$ (Eq. (\ref{alpha})) as a function of the
strong phase difference $\delta$, solid curve $\delta_T=\pi/12$,  and
dashed curve $\delta_T =0$.}
\label{fi:alpha}
\end{figure}
\begin{figure}
\epsfxsize=12cm
\centerline{\epsfbox{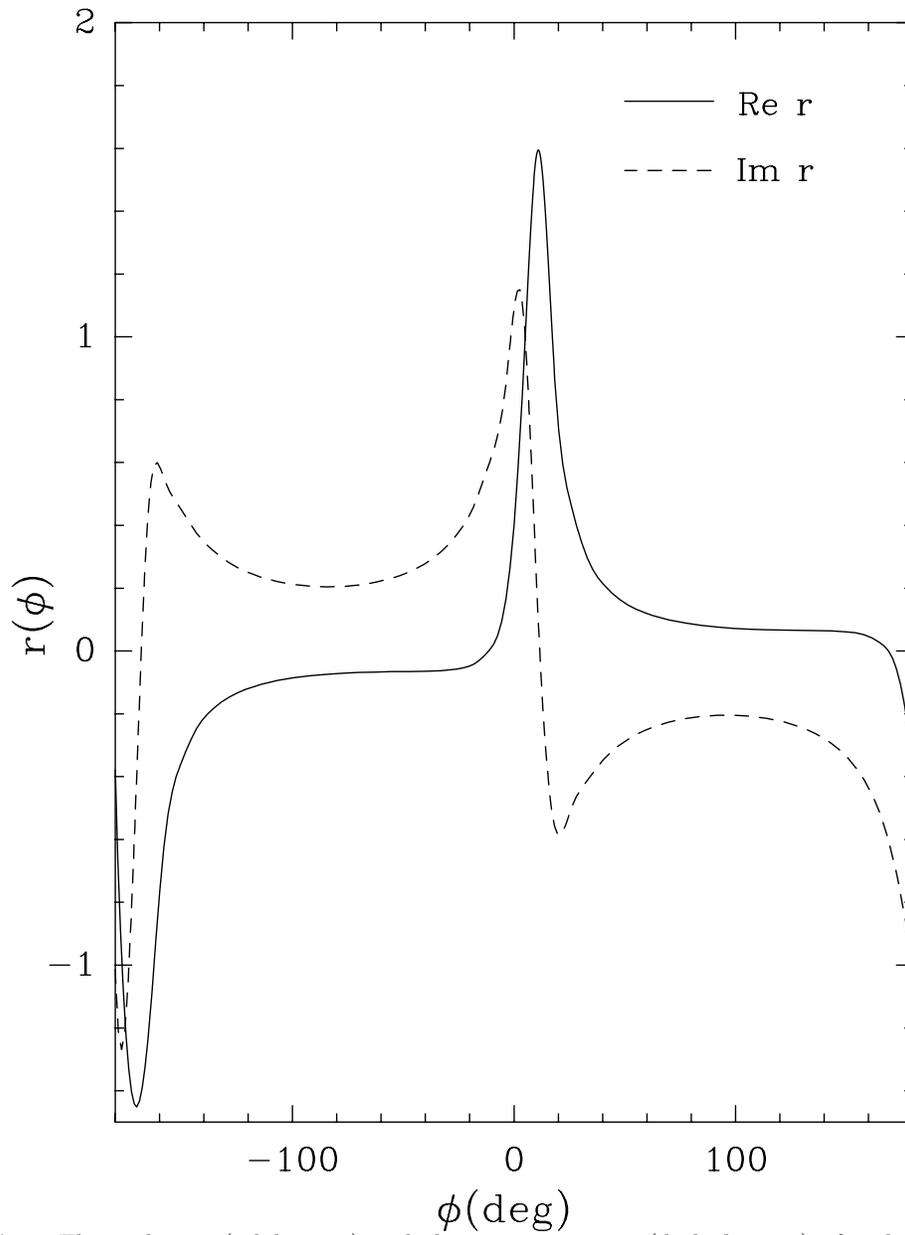}}
\caption{The real part (solid curve) and the imaginary part (dashed curve) of
$r$ defined in (\ref{rpik})-(\ref{rpik1}), as functions of the strong phase
difference $\phi$, for $\phi_P=\pi/6$.}
\label{fi:rpik}
\end{figure}
\begin{figure}
\epsfxsize=12cm
\centerline{\epsfbox{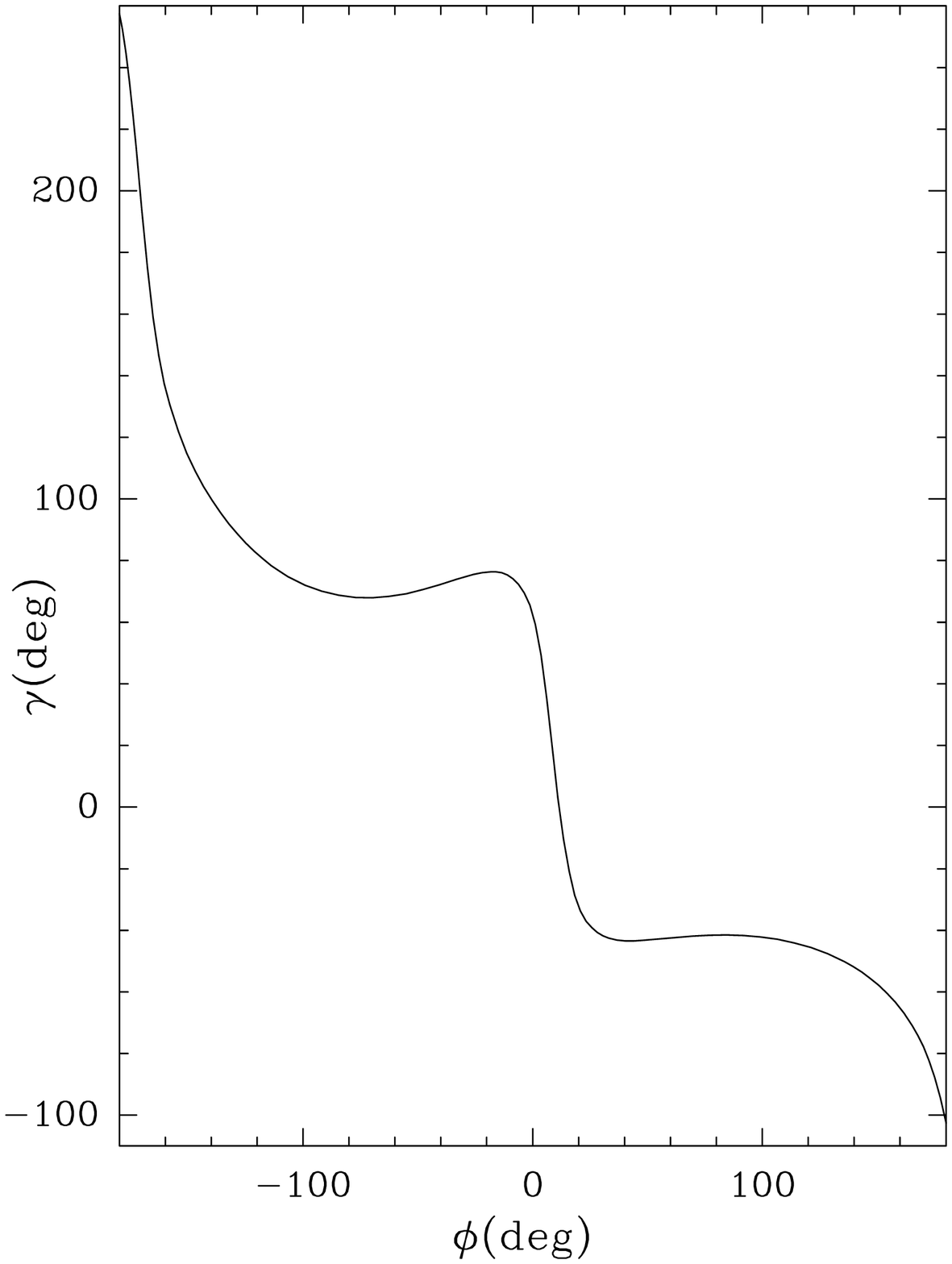}}
\caption{The weak phase $\gamma=\mbox{Arg}\, (r/\epsilon_{3/2}+\delta_{EW})$, 
as a function of the strong phase difference $\phi$, for $\phi_P =\pi/6$.}
\label{fi:gama}
\end{figure}


\begin{thebibliography}{99}

\bibitem{CaMi} I. Caprini, L. Micu, and C. Bourrely, Phys. Rev. D {\bf 60},
074016 (1999).

\bibitem{KaWi} G. K\"allen and A.S. Wightman, Dan. Vid. Selsk.
Mat-fys. Skr. {\bf 1}, no 6 (1958).

\bibitem{LSZ} H. Lehmann, K. Symanzik, and W. Zimmermann, Nuovo Cimento,
{\bf 1}, 205 (1956); {\bf 2}, 425 (1957).

\bibitem{Bart} G. Barton, {\it Introduction to Dispersion Techniques in Field
 Theory}, (Benjamin, New York, 1965).

\bibitem{GoTr} M.L. Goldberger and S.B. Treiman, Phys. Rev.
{\bf 110}, 1178 (1958); {\bf 111}, 354 (1958).

\bibitem{Omnes} R. Omn\`es, Nuovo Cimento {\bf 8}, 316 (1958).

\bibitem{Dono1} J.F. Donoghue {\it et al.}, Phys. Rev. Lett. {\bf 77}, 2178
(1996).

\bibitem{Coll} P.D.B. Collins, {\it Introduction to Regge Theory and
High Energy Physics} (Cambridge University Press, Cambridge, England, 1977).

\bibitem{Dele} D. Del\'epine, J.-M. G\'erard, J. Pestiau, and J. Weyers, Phys.
Lett. B {\bf 429}, 106 (1998).


\bibitem{Gron} M. Gronau, O.F. Hernandez, D. London, and
J.L. Rosner, Phys. Rev. D{\bf 50}, 4529 (1994).

\bibitem{BlokG} B. Blok, M. Gronau and J.L. Rosner, Phys. Rev. Lett.
{\bf 78}, 3999 (1997).

\bibitem{BeBu} M. Beneke, G. Buchalla, M. Neubert, and C.T. Sachrajda,
Phys. Rev. Lett. {\bf 83}, 1914 (1999).

\bibitem{Neub} M. Neubert, JHEP 9902:014 (1999).

\bibitem{Falk} A.F. Falk, A.L. Kagan, Y. Nir, and A.A. Petrov,
Phys. Rev. D {\bf 57}, 4290 (1998).

\end{thebibliography}
\end{document}